\documentclass[a4paper,12pt]{article}
\usepackage{amssymb}
\usepackage{latexsym}
\usepackage[dvips]{graphicx}
\newcommand{\be}{\begin{equation}}
\newcommand{\ee}{\end{equation}}
\newcommand{\bea}{\begin{eqnarray}}
\newcommand{\nn}{\nonumber}
\newcommand{\eea}{\end{eqnarray}}
\begin{document}

\begin{titlepage}
\begin{flushright}
hep-th/0108147\\
UA/NPPS-10-2001
\end{flushright}
\begin{centering}
\vspace{.8in}
{\large {\bf Two-Dimensional Dilatonic Black Holes\\ and\\ Hawking Radiation}} \\

\vspace{.5in}
{\bf Elias C. Vagenas\footnote{hvagenas@cc.uoa.gr}} \\

\vspace{0.3in}
University of Athens\\ Physics Department\\
Nuclear and Particle Physics Section\\
Panepistimioupolis, Ilisia 157 71\\ Athens, Greece\\
\end{centering}

\vspace{1in}
\begin{abstract}
\noindent Hawking radiation emanating from  two-dimensional
charged and uncharged dilatonic black holes - dimensionally reduced from
 $(2+1)$ spinning and spinless, respectively, BTZ black holes - 
is viewed as a tunnelling process. Two
dimensional dilatonic black holes (AdS(2) included) are treated as
dynamical backgrounds in contrast to the standard methodology
where the background geometry is fixed when evaluating Hawking
radiation. This modification to the geometry gives rise to a
nonthermal part in the radiation spectrum. Nonzero temperature of
the extremal two-dimensional charged black hole is found.
 The Bekenstein-Hawking area formula is easily derived for these
 dynamical geometries.
 \\
\\
\\
\\
\\
\\
\\
\end{abstract}
\end{titlepage}
\section*{Introduction}
In 1992 Ba$\tilde{n}$ados, Teitelboim and Zanelli
(BTZ) \cite{banados1,banados2} showed that ($2+1$)-dimensional
gravity has a black hole
 solution. This black hole is described by two parameters,
its mass $M$ and its angular momentum (spin) $J$. It is locally
anti-de-Sitter space and thus it differs from Schwarzschild and
Kerr solutions in that it is asymptotically anti-de-Sitter instead
of flat. Additionally it has no curvature singularity at the
origin.
 AdS black holes, which are members of the two-parametric family
of BTZ black holes, play a central role in AdS/CFT conjecture
 \cite{sen,maldacena1} and also in brane-world scenarios
 \cite{RS1,RS2}.
Specifically AdS(2) black hole is most interesting in the context
of string theory and black hole physics
\cite{strominger1,strominger2}. 
\par\noindent In this paper motivated by this
recent interest in  two-dimensional black hole backgrounds we
treat the two-dimensional charged and uncharged dilatonic black holes (including AdS(2)) as
radiating sources. \par \noindent Concerning the quantum process
called Hawking effect \cite{hawking1} much work has been done
using a fixed background during the emission process. The idea of
Keski-Vakkuri, Kraus and Wilczek (KKW) \cite{KKW1,KKW2,KKW3,KKW4}
is to view the black hole background as dynamical by treating the
Hawking radiation as a tunnelling process. The energy conservation
is the key to this description. The total (ADM) mass is kept fixed
while the mass of the two-dimensional dilatonic black hole decreases due to the emitted
radiation. The effect of this modification gives rise to
additional terms in the formulae concerning the known results for
the two-dimensional charged and uncharged black holes; these additions are analogous to those found
in \cite{correction1,correction2,correction3,correction4} for the
respective geometries; a nonthermal partner to the thermal spectrum of the
Hawking radiation shows up. We explore the consequences to vacuum
states of two-dimensional  black holes (extremal black holes) since black
holes are in general regarded as highly excited states. The
extremality of the two-dimensional charged black
holes is now shifted since the charge $J$ of the 
charged black hole is approached by the mass $M$ earlier. This
alteration produces a non -``frozen" extremal two-dimensional
charged black hole characterized by a constant Hawking temperature
$\,T^{ext}_{H}\neq0\,$. KKW method provides an easy way to
derive the entropy of the two-dimensional charged and uncharged black holes.
\par \noindent
In section 1 we make a short review of the two-dimensional
charged and uncharged dilatonic  black holes and their
properties. We present for the two-dimensional charged  black
hole expressions for its temperature, area and entropy. The
extremal two-dimensional charged black hole is derived and
its zero temperature is verified.  In section 2
we implement the KKW method. Using the imaginary part of the
action of an outgoing positive-energy particle the temperature of
the two-dimensional charged black hole is evaluated and its
dependence on the energy of the emitted massless particle is
explicitly shown. This modified temperature due to the
self-gravitation effect leads to a nonthermal spectrum. The
extremal (vacuum state) two-dimensional charged black hole is
no more ``frozen''. Its temperature is nonzero. Corresponding
results for the two-dimensional uncharged black holes and for the 
AdS(2) spacetime are deduced. In section 3 we calculate the entropy of two-dimensional
charged and uncharged black holes.  Finally in section 4 we
summarize our results and give our conclusions.
\section{($1+1$) Dilatonic Black Holes}
The black hole solutions of Ba$\tilde{n}$ados, Teitelboim and
Zanelli in ($2+1$) spacetime dimensions are derived from a three
dimensional theory of gravity: \be S=\int dx^{3}
\sqrt{-g}\,({}^{{\small(3)}} R+2\Lambda) \ee with a negative
cosmological constant ($\Lambda>0$). The corresponding line
element is: \be ds^2 =-\left(-M+\Lambda r^2 +\frac{J^2}{4 r^2}
\right)dt^2 +\frac{dr^2}{\left(-M+\Lambda r^2 +\frac{J^2}{4 r^2}
\right)}+r^2\left(d\theta -\frac{J}{2r^2}dt\right)^2. \label{btzmetric}\ee 
There are many  ways to reduce the three dimensional BTZ black hole
solutions to the two dimensional charged and uncharged dilatonic black holes \cite{ortiz,lowe}. 
The Kaluza-Klein reduction of the ($2+1$)-dimensional metric (\ref{btzmetric}) yields a  
two-dimensional line element:
 \be ds^2 =-g(r)dt^2 +g(r)^{-1}dr^2
\label{metric1}\ee where \be g(r)=\left(-M+\Lambda r^2
+\frac{J^2}{4 r^2}\right)\label{metric2}
 \ee
with $M$ the ADM mass, $J$ the charge 
 of the two-dimensional charged black hole, a U($1$) gauge field:
\be
A_{t}=-\frac{J}{2r^{2}}
\ee
and a dilaton field:
\be
\Phi= r.
\ee
\par \noindent
For the positive mass black hole spectrum with charge ($J\neq 0$),
the line element (\ref{metric1}) has two horizons: \be
r^{2}_{\pm}=\frac{M\pm\sqrt{M^2 - \Lambda J^2} }{2\Lambda}
\label{horizon1} \ee with $r_{+}$, $r_{-}$ the outer and inner
horizon respectively. The area $\mathcal{A}_H$ and Hawking
temperature $T_H$ of the event (outer) horizon are
\cite{kumar1,kumar2}: \bea \mathcal{A}_H
&=&2\pi\left(\frac{M+\sqrt{M^2 +\Lambda J^2
}}{2\Lambda}\right)^{1/2}\nn\\
&=&2\pi r_{+} \\
T_H &=&
\frac{\sqrt{2\Lambda}}{2\pi}\frac{\sqrt{M^2-\Lambda J^2}}
{(M+\sqrt{M^2 -\Lambda J^2})^{1/2}}\nn\\
&=&\frac{\Lambda}{2\pi}\left(\frac{r_{+}^{2}-r_{-}^{2}}{r_{+}}\right).
\label{temp1}\eea 
The entropy of the two-dimensional charged black hole is:
\be S_{bh}=4\pi r_{+} \label{entropy1}\ee and if we reinstate the
Planck units $8\hbar G =1 $ we get: \be
S_{bh}=\frac{1}{4\hbar G} \mathcal{A}_H=S_{BH} \ee which is the
well-known Bekenstein-Hawking area formula for the entropy ($S_{BH}$)
\cite{bekenstein1,bekenstein2,hawking3} (or
\cite{strominger3} by counting excited states).
\par \noindent
Concerning the extremality of the two-dimensional charged black hole: \be
M=\sqrt{\Lambda}J \ee the inner and outer horizon coincide
($r_{+}=r_{-}$). This two-dimensional charged black hole can be viewed as the vacuum
state of the positive mass spectrum of two-dimensional charged black holes
which saturates the bound: \be M^2 - \Lambda J^2 \geq 0
\Leftrightarrow M \geq \sqrt{\Lambda}|J| \ee imposed in
(\ref{horizon1}). Obviously the extremal two-dimensional charged black hole
has zero temperature ($T_{H}^{ext}=0$).
\par \noindent
The two-dimensional uncharged black hole may be obtained by a similar Kaluza-Klein 
reduction for a covariantly constant electric field and the resulting metric is:
 \be ds^2 =-\left(-M+\Lambda r^2 \right) dt^2
+\left(-M+\Lambda r^2\right)^{-1}dr^2 \ee which has an horizon at:
\be r_{H}=\sqrt{\frac{M}{\Lambda}} \label{horizon2} \ee and is
similar to Schwarzschild black hole with the important difference
that it is not asymptotically flat but it has constant negative
curvature. The temperature of the two-dimensional uncharged black hole is \cite{kim}: \be
T_H = \frac{\sqrt{2\Lambda}}{2\pi}\; M^{1/2}\label{temp2}. \ee
\par \noindent Two-dimensional charged black holes with $M < \sqrt{\Lambda}|J|$
  are discarded since they contain a naked singularity and for the
 same reason two-dimensional uncharged black holes with $M<0$ have not been
 treated above. The only exception is the two-dimensional uncharged black hole with $M=-1$
 and $J=0$ which is the ordinary anti-de Sitter spacetime
 (AdS(2) black hole).
\section{KKW Methodology}
In order to apply the idea of Keski-Vakkuri, Kraus and Wilczek
(KKW) \cite{KKW1,KKW2,KKW3,KKW4} to the two-dimensional charged black hole
 (\ref{metric1}-\ref{metric2}) we have
to make a coordinate transformation. We choose the
Painlev$\acute{e}$ coordinates \cite{painleve} (utilized for black
hole backgrounds recently in \cite{wilczek}) which are
non-singular on the outer horizon ($r_{+}$). Thus we will be able
to deal with phenomena whose main contributions come from the
outer horizon. We introduce the time coordinate $\tau_P$ by
imposing the ans\"{a}tz:
\be \sqrt{g(r)}\,dt=\sqrt{g(r)}\,d\tau_{P}- \sqrt{1+M-\Lambda r^2
-\frac{J^2}{4 r^2}}\,\frac{dr^2}{\sqrt{g(r)}}. \ee
The line element (\ref{metric1}-\ref{metric2}) is now written as :
\be ds^2 = -\left(-M+\Lambda r^2 +\frac{J^2}{4
r^2}\right)d\tau_P^2
 + 2 \sqrt{1+M-\Lambda r^2-\frac{J^2}{4 r^2}}\,d\tau_P dr + dr^2.
\label{metric3}
 \ee It is obvious from the
above expression that there is no singularity at the points  $r_+$
and $r_-$. The null ($ds^2 =0$) geodesics followed by a
massless particle are: \be \dot{r}\equiv\frac{dr}{d\tau_P}=\pm 1-
\sqrt{1+M-\Lambda r^2-\frac{J^2}{4 r^2}} \ee where the signs $+$
and $-$ correspond to the outgoing and ingoing geodesics,
respectively, under the assumption that $\tau_P$ increases towards
future.\par \noindent
 We fix the total ADM mass and we let the
mass $M$ of the two-dimensional charged black hole vary. If a shell of energy (mass)
$\omega$ is radiated outwards the outer horizon then the two-dimensional charged black
hole mass will be reduced to $M-\omega$ and the shell of energy
will travel on the modified geodesics: \be \dot{r}=1-
\sqrt{1+(M-\omega)-\Lambda r^2-\frac{J^2}{4 r^2}}
\label{geodesic}\ee produced by the modified line element: \be
ds^2 = -\left(-(M-\omega)+\Lambda r^2 +\frac{J^2}{4
r^2}\right)d\tau_P^2
 + 2 \sqrt{1+(M-\omega)-\Lambda r^2-\frac{J^2}{4 r^2}}\,d\tau_P dr +
 dr^2 .
 \label{metric4}
\ee
\par\noindent
It is known that the emission rate from a radiating source can be
expressed in terms of the imaginary part of the action for an
outgoing positive-energy particle as: \be \Gamma=e^{-2ImS}
\label{gamma}\ee but also in terms of the temperature and the
entropy of the radiating source which in our case will be a two-dimensional charged 
black hole: \be \Gamma=e^{-\beta \omega}=e^{+\Delta
S_{\hspace{0.05cm}bh}} \label{rate1} \ee where $\beta$ is the
inverse temperature of the two-dimensional charged black hole and $\Delta
S_{\hspace{0.05cm}bh}$ is the change the entropy of the two-dimensional charged black
hole before and after the emission of the shell of energy $\omega$
(outgoing massless particle). It is clear that if we evaluate the
action then we will know the temperature and/or the change in the
entropy of the two-dimensional charged black hole. We therefore evaluate the imaginary
part of the action for an outgoing positive-energy particle which
crosses the event horizon outwards from:
 \be
r_{in}=r_{+}(M, J)=\left(\frac{M+\sqrt{M^2 - \Lambda J^2}
}{2\Lambda}\right)^{1/2}
\hspace{1.5cm}\ee to 
\be r_{out}=r_{+}(M-\omega, J)=\left(\frac{(M-\omega)+\sqrt{(M-\omega)^2 - \Lambda J^2}
}{2\Lambda}\right)^{1/2}. \ee The imaginary part of the action is: \be
ImS=Im\int_{r_{in}}^{r_{out}}p_{r}dr=Im\int^{r_{out}}_{r_{in}}
\int_{0}^{p_{r}}dp'_{r}dr. \ee We make the transition from the
momentum variable to the energy variable using Hamilton's equation
$\dot{r}=\frac{dH}{dp_{r}}$  and equation (\ref{geodesic}). The
result is: \be
ImS=Im\int^{r_{out}}_{r_{in}}\int^{\omega}_{0}\frac{(-d\omega')dr}
{1-\sqrt{1+(M-\omega')-\Lambda r^2 -\frac{J^2}{4 r^2}}} \ee where
the minus sign is due to the Hamiltonian being equal to the
modified mass $H=M-\omega$. This is not disturbing since
$r_{in}>r_{out}$. After some calculations (involving contour
integration into the lower half of $\omega'$ plane) we get: \be
ImS=2\pi\Bigg[\left(\frac{M+\sqrt{M^2 - \Lambda J^2}
}{2\Lambda}\right)^{1/2}-\left(\frac{(M-\omega)+\sqrt{(M-\omega)^2
- \Lambda J^2} }{2\Lambda}\right)^{1/2}\Bigg]. \ee Apparently the
emission rate depends not only on the mass $M$ and charge
 $J$ of the two-dimensional charged black hole but also on the energy
$\omega$ of the emitted massless particle: \bea
\Gamma(\omega,M,J)=e^{-2ImS}\hspace{11cm}&&\nn\\
= exp\left[ 4\pi\left(\sqrt{\frac{(M-\omega)+\sqrt{(M-\omega)^2 -
\Lambda J^2} }{2\Lambda}}-\sqrt{\frac{M+\sqrt{M^2 - \Lambda J^2}
}{2\Lambda}}\;\right)\right].&& \label{temp2} \eea Comparing
(\ref{rate1}) and (\ref{temp2}) we deduce that the modified
temperature (due to the self-gravitation) of the two-dimensional charged black hole is:
 \be T(\omega)= \frac{\omega}{4\pi}\left[ \left(\frac{M+\sqrt{M^2
- \Lambda J^2}
}{2\Lambda}\right)^{1/2}-\left(\frac{(M-\omega)+\sqrt{(M-\omega)^2
- \Lambda J^2} }{2\Lambda}\right)^{1/2}\right]^{-1}
\label{temp3}.\ee We see that there are modifications to the
result previously mentioned (\ref{temp1}) for a fixed two-dimensional charged black
hole background. The temperature of the two-dimensional charged black hole is no
longer the Hawking temperature $T_H$. 
\\
\subsection{ ($1+1$) Charged Black Hole}
\par\noindent
Concerning the extremal (vacuum solution) two-dimensional charged black hole
the condition for extremality now satisfied will be: \be
M-\omega=\sqrt{\Lambda}J \;. \ee This modification indicates that
the mass $M$ of the two-dimensional charged black hole cannot be less than
the charge $J$ (since
$\omega=M-\sqrt{\Lambda}J>0$) and the temperature of the extremal
two-dimensional charged black hole will not be zero but: \be
T^{ext}_{H}=\frac{\sqrt{2\Lambda}}{4\pi}\;
\frac{(M-\sqrt{\Lambda}J)}{(M+\sqrt{M^2 -\Lambda
J})^{1/2}-(\sqrt{\Lambda}J)^{1/2}}\;. \ee
\\
\subsection{($1+1$) Uncharged Black Hole}
The modified temperature (\ref{temp3}) for the two-dimensional uncharged black
hole becomes: \be T(\omega)= \frac{\omega}{4\pi}\left[
\sqrt{\frac{M}{\Lambda}}-\sqrt{\frac{M-\omega}{\Lambda}}\right]
\label{temp4} \ee and which in first order in $\omega$ is: \be T_H
=\frac{\sqrt{\Lambda}}{2\pi}\;M^{1/2} \ee  in agreement to what
was shown in \cite{correction2}.
\par \noindent 
For the case $M=-1$ which may be
recognized as the ordinary anti-de-Sitter space (AdS(2) spacetime)
the modified temperature is: \be T(\omega) =\frac{\omega
\sqrt{\Lambda}}{4\pi}\left(1-\sqrt{1-\omega}\right)^{-1} \ee
which, to first order in $\omega$, gives: \be
T_{H}=\frac{\sqrt{\Lambda}}{2\pi}\; . \ee
\section{Entropy Calculation via KKW Method}
It is obvious that we can have a short and direct derivation of
the entropy of two-dimensional charged black hole up to a constant using equations
(\ref{gamma}) and (\ref{rate1}) where: \be \Delta
S_{\hspace{0.05cm}bh} = S_{bh}(M-\omega, J)- S_{bh}(M, J). \ee Indeed, if we
combine (\ref{gamma}), (\ref{rate1}) and (\ref{temp3}) we get:
\be
S_{bh}(\omega, M, J)=4\pi\left[\frac{(M-\omega)+\sqrt{(M-\omega)^2 -\Lambda J^2
}}{2\Lambda}\right]^{1/2} + S_0 
\ee 
where $S_0$ is the arbitrary
constant and which to first order in $\omega$ will give the known expression (\ref{entropy1}) 
up to a constant for the entropy
of the two-dimensional charged black hole: 
\be
S_{bh}(M, J)=4\pi\left[\frac{M+\sqrt{M^2 -\Lambda J^2
}}{2\Lambda}\right]^{1/2} + S_0 .
\ee
\par \noindent
For the two-dimensional uncharged black hole the entropy is: \be
S_{bh}(M)=4\pi\sqrt{\frac{M-\omega}{\Lambda}}+S_0' \ee where $S_0 '$ is also
an arbitrary constant and which will give to first order in $\omega$:
\be
S_{bh}(M)=4\pi\sqrt{\frac{M}{\Lambda}}+S_0'.
\ee
 If we adopt the conjecture that the entropy
of an excited state tends to the entropy of its vacuum state
\cite{mayo} then the additive constants are set to zero : 
\be S_0 = 0\hspace{0.5cm}\mbox{and}\hspace{0.5cm}
S_0'=0 \; .\ee 
\section*{Discussion}
In this work, we have viewed the Hawking radiation as a quantum
tunnelling process. The self-gravitation of the radiation was
included and this treatment introduced a nonthermal part for the
radiation spectrum of the two-dimensional charged and uncharged dilatonic black holes. 
The temperature of the two-dimensional charged black hole is no more the Hawking
temperature and the ``greybody factor" showing up declares
explicitly the dependence of the temperature on the emitted
particle's energy $\omega$. The leading term in $\omega$ gives the
thermal Boltzmann factor while the higher order terms represent
corrections emanating from the response of the background geometry
to the emission of a quantum. The extremal two-dimensional
charged black hole is no more ``frozen" but it carries a
background Hawking temperature $\,T^{ext}_{H}\neq 0\,$ ensuring the
validity of the third law of black hole thermodynamics
\cite{liberati}. Therefore it is obvious that we again have a
strong evidence to believe that black holes constitute excited
states while the extremal black holes correspond to ground
(vacuum) states.
\par
The above-mentioned treatment for incorporating the effects of the
emission of a shell of energy for the case of two-dimensional
uncharged dilatonic black holes (including AdS(2) black hole) yields the
corresponding modified temperature with the leading term in
$\omega$ again being the thermal Boltzmann factor.
 \par \noindent
Additionally the imaginary part of the action of the outgoing
positive-energy particle is linear in the change of the entropy.
We derive in a short and direct way the modified entropy for the two-dimensional 
charged and uncharged dilatonic black holes due to the specific modelling of the 
self-gravitation effect.  A welcomed but not unexpected result is that
the expressions for the entropy of  the two-dimensional dilatonic black holes obtained to first order in $\omega$ are in
agreement with those for the corresponding fixed two-dimensional  black hole backgrounds. 


\section*{Acknowledgements}
The author would like to thank Ass. Professor T. Christodoulakis
 and Ass. Professor G.A. Diamandis for useful discussions. 
The author also thanks the referee 
 for his comments which helped to clarify points in the manuscript. This work is
financially supported in part by the University of Athens' Special
Account for the Research.



 \end{document}